\documentclass{llncs}

\newcommand\rea{\mbox{$\mathbb{R}$}}

\newcommand{\marginlabel}[1]%
{\mbox{}\marginpar{\it{\raggedleft\hspace{0pt}#1}}}
\newcommand\set[1]{\left\{#1\right\}} 

\newcommand\ket[1]{| #1 \rangle}
\newcommand\bra[1]{\langle #1 |}

\newcommand\av{\mbox{\bf{\bf E}}}

\newcommand\mcE{\mathcal E}
\newcommand\mcF{\mathcal F}
\newcommand\mcG{\mathcal G}

\newcommand\mcP{\mathcal{P}}

\newcommand\mcX{\mathcal{X}}
\newcommand\mcY{\mathcal{Y}}
\newcommand\mcZ{\mathcal{Z}}
\newcommand\mcM{\mathcal{M}}

\newcommand{\bw}{{\bf{w}}}

\newlength{\pgmtab}  
\newcommand{\sD}{{\mathsf{D}}}
\newcommand{\sR}{{\mathsf{R}}}
\newcommand{\sQ}{{\mathsf{Q}}}

\newcommand{\pub}{{\mathsf{pub}}}
\newcommand{\vc}{{\mathsf{VC}}}
\newcommand{\alice}{{\mathsf{Alice}}}
\newcommand{\bob}{{\mathsf{Bob}}}

\newcommand{\SMP}{{\mathsf{SMP}}}

\newcommand{\DISJ}{{\mathsf{DISJ}}}
\newcommand{\IP}{{\mathsf{IP}}}
\newcommand{\GT}{{\mathsf{GT}}}

\newcommand{\defeq}{\stackrel{\mathsf{def}}{=}}
\newcommand{\norm}[1]{\left\| #1 \right\|}
\newcommand{\Tr}{{\mathsf{Tr}}}
\newcommand{\Good}{{\mathsf{GOOD}}}
\newcommand{\rec}{{\mathsf{rec}}}

\newcommand{\pg}{{\mathcal{P}_\gamma}}

\newcommand\llra\longleftrightarrow

\usepackage{amsmath}
\usepackage{xspace}
\usepackage{amssymb}
\newenvironment{proofof}[1]{\noindent{\bf Proof of #1:}}{\qed}

\newtheorem{fact}{Fact}    

\newsavebox{\fmbox}
\newsavebox{\algobox}

 {
    \begin{enumerate}}{\end{enumerate}}

\title{New bounds on classical and  quantum one-way communication complexity}

\author{Rahul Jain\inst{1}\thanks{School of Computer Science, and
Institute for Quantum Computing,
University of Waterloo, 200 University Ave.\ W., Waterloo, ON N2L 3G1,
Canada. Research supported in part by ARO/NSA USA.
}
\and 
Shengyu Zhang\inst{2}
\thanks{Computer Science Department and Institute for Quantum
Computing, California Institute of Technology, 1200 E California Bl, IQI, MC 107-81, Pasadena, CA 91125, USA. This work was supported by the National Science Foundation under grant PHY-0456720 and the Army Research Office under
grant W911NF-05-1-0294 through Institute for Quantum Information at California Institute of Technology.}   
}
\institute{University of Waterloo \email{\quad rjain@cs.uwaterloo.ca} 
\and California Institute of Technology \email{\quad shengyu@caltech.edu}
}

\begin{document}
\begin{titlepage}
\maketitle
\thispagestyle{empty}

\abstract {
In this paper we provide new bounds on classical and quantum
distributional communication complexity in the  two-party, one-way
model of communication.

\vspace{0.2cm}
In the classical one-way model, our bound extends the well known upper
bound of Kremer, Nisan and Ron~\cite{KremerNR95} to include
non-product distributions. Let $\epsilon \in (0,1/2)$ be a
constant. We show that for a boolean function $f : \mcX \times \mcY
\rightarrow \{0,1\}$ and a non-product distribution $\mu$ on $\mcX
\times \mcY$, $$ \sD_{\epsilon}^{1,\mu}(f) = O((I(X:Y)+1) \cdot
\vc(f)),$$
where $\sD_{\epsilon}^{1,\mu}(f)$ represents the one-way distributional
communication complexity of $f$ with error at most $\epsilon$ under
$\mu$; $\vc(f)$ represents the {\em Vapnik-Chervonenkis} dimension of $f$ and
$I(X:Y)$ represents the mutual information, under $\mu$, between
the random inputs of the two parties. For a non-boolean
function $f : \mcX \times \mcY \rightarrow \set{1, \ldots, k}$ ($k
\geq 2$ an integer), we show a similar upper bound on $\sD_{\epsilon}^{1,\mu}(f)$
in terms of $k, I(X:Y)$ and the {\em pseudo-dimension}
of $f' \defeq \frac{f}{k}$, a generalization of the $\vc$-dimension for non-boolean functions.

\vspace{0.2cm}
In the quantum one-way model we provide a lower bound on the
distributional communication complexity, under product distributions,
of a function $f$, in terms the well studied complexity measure of $f$
referred to as the {\em rectangle bound} or the {\em corruption bound}
of $f$. We show for a non-boolean total function $f : \mcX \times \mcY
\rightarrow \mcZ$ and a product distribution $\mu$ on $\mcX
\times \mcY$,
$$ \sQ_{\epsilon^3/8}^{1,\mu}(f) = \Omega(\rec_\epsilon^{1,\mu}(f)), $$
where $\sQ_{\epsilon^3/8}^{1,\mu}(f)$ represents the quantum one-way
distributional communication complexity of $f$ with error at most
$\epsilon^3/8$ under $\mu$ and $\rec_\epsilon^{1,\mu}(f)$ represents the
one-way rectangle bound of $f$ with error at most $\epsilon$ under
$\mu$. Similarly for a non-boolean partial function $f : \mcX \times \mcY
\rightarrow \mcZ \cup \set{*}$ and a product distribution $\mu$ on $\mcX
\times \mcY$, we show,
$$ \sQ_{\epsilon^6/(2\cdot 15^4)}^{1,\mu}(f) = \Omega(\rec_\epsilon^{1,\mu}(f)). $$
}
\end{titlepage}

\section{Introduction}

Communication complexity studies the minimum amount of communication
that two or more parties need to compute a given function or a
relation of their inputs. Since its inception in the seminal paper by
Yao~\cite{Yao79}, communication complexity has been an important and
widely studied research area. This is the case both because of the interesting
and intriguing mathematics involved in its study, and also because of
the fundamental connections it bears with many other areas in theoretical computer
science, such as data structures, streaming algorithms, circuit lower
bounds, decision tree complexity, VLSI designs, etc.

Different models of communication have been proposed and studied. In
the basic and standard \textit{two-party interactive} model, two
parties say $\alice$ and $\bob$, each receive an input say $x \in
\mcX$ and $y \in \mcY$, respectively. They interact with each other
possibly communicating  several messages in order to jointly compute, say a given
function $f(x,y)$ of their inputs. If only one message is allowed, say from $\alice$
to $\bob$, and $\bob$ outputs $f(x,y)$ without any further interaction
with $\alice$, then the model is said to be
\textit{one-way}. Though seemingly simple, this model has numerous nontrivial questions as well as applications to other areas such as lower bounds of streaming algorithms, see for example~\cite{Mut05}. Other models like the {\em
Simultaneous message passing} ($\SMP$) model, and {\em multi-party} models
are also studied. We refer readers to the
textbook~\cite{KushilevitzN97} for a comprehensive introduction to the
field of classical communication complexity.
In 1993, Yao~\cite{Yao93} introduced {\em quantum} communication
complexity and since then it has also become a very active and vibrant
area of research. In the quantum communication models, the
parties are allowed to use quantum computers to process their inputs
and to use quantum channels to send messages.

In this paper we are primarily concerned with the one-way model and we
assume that the single message is always, say from $\alice$ to $\bob$.
Let us first briefly discuss a few classical models. In the
deterministic one-way model, the parties act in a deterministic
fashion, and compute $f$ correctly on all input pairs $(x,y)$. The
minimum communication required for accomplishing this is called as the
{\em deterministic complexity} of $f$ and is denoted by $\sD^1(f)$.
Allowing the parties to use randomness and allowing
them to err on their inputs with a small non-zero probability, often
results in considerable savings in communication. The communication of
the best public-coin one-way protocol that has error at most
$\epsilon$ on all inputs, is referred to as the {\em one-way
public-coin randomized} communication complexity of $f$ and is denoted
by $\sR^{1,\pub}_\epsilon(f)$. Similarly we can define the {\em
one-way private-coin randomized} communication complexity of $f$,
denoted by $\sR^{1}_\epsilon(f)$ and in the quantum model, the {\em
one-way quantum} communication complexity of $f$, denoted by
$\sQ^{1}_\epsilon(f)$. Please refer to Sec.~\ref{sec:commprelim}
for explicit definitions.  When the subscript is omitted, $\epsilon$ is assumed to be
$1/3$.

Sometimes the requirement on communication protocols is less stringent
and it is only required that the average error, under a given
distribution $\mu$ on the inputs, is small. The communication of the
best one-way classical protocol that has average error at most
$\epsilon$ under $\mu$, is referred to as the {\em one-way
distributional} communication complexity of $f$ and is denoted by
$\sD^{1,\mu}_\epsilon(f)$.  
We can define the one-way distributional quantum communication
complexity $\sQ^{1,\mu}_\epsilon(f)$ in a similar way.  A useful
connection between the public-coin randomized and distributional
communication complexities via the Yao's Principal~\cite{Yao77} states
that for a given $\epsilon \in 
(0,1/2)$, $\sR_\epsilon^{1, \pub}(f) = \max_\mu
\sD_\epsilon^{1, \mu}(f)$. A distribution $\mu$, that
achieves the maximum in Yao's 
Principal, that is for which $\sR_\epsilon^{1, \pub}(f) =
\sD_\epsilon^{1, \mu}(f)$, is referred to as a {\em hard}
distribution for $f$. This principal also holds in many other models
and allows for a good handle on the public-coin randomized complexity
in scenarios where the distributional complexity is much easier to
understand.  Often, the distributional complexity when the inputs of
$\alice$ and $\bob$ are drawn independently from a product
distribution, is easier to understand.  Nonetheless, often as is
the case with several important functions like Set Disjointness
($\DISJ$) and Inner Product ($\IP$), the maximum in Yao's Principal,
in the one-way model, occurs for a product distribution, and hence it
paves the way for understanding the public-coin randomized complexity.

Let us now discuss our first main result which is in the classical
one-way model.  We ask the reader to refer to
Sec.~\ref{sec:prelim} for the definitions of various quantities
involved in the discussion below.

\subsection{Classical upper bound}
For a boolean function $f: \mcX \times \mcY
\rightarrow \{0,1\}$, its {\em Vapnik-Chervonenkis (VC) dimension}, denoted by
$\vc(f)$, is an important complexity measure, widely studied specially in the contexts of {\em computational
learning theory}. Kremer, Nisan and Ron~\cite[Thm.~3.2]{KremerNR95} found a
beautiful connection between the distributional complexity of $f$ under
product distributions on $\mcX \times
\mcY$, and $\vc(f)$, as follows.

\begin{theorem}[\cite{KremerNR95}]
\label{thm:vc-dim}
Let~$f : \mcX \times \mcY \rightarrow \{0,1\}$ be a boolean function
and let~$\epsilon \in (0, 1/2)$ be a constant. Let $\mu$ be a {\em product} distribution
on $\mcX \times \mcY$. There is a universal constant~$\kappa$ such
that,
\begin{equation}
\label{eq:kremer}
    \sD^{1,\mu}_{\epsilon}(f) 
    \quad \leq \quad \kappa \cdot \frac{1}{\epsilon}
               \log\frac{1}{\epsilon} \cdot \vc(f).
\end{equation}

\end{theorem}
Note that such a relation cannot hold for non-product distributions
$\mu$ since otherwise it would translate, via the Yao's Principal,
into $\sR^{1, \pub}_\epsilon(f) = O(\vc(f))$, for all boolean
$f$. This is not true as is exhibited by several functions for example
the Greater Than ($\GT_n$) function, in which 
$\alice$ and $\bob$ need to determine which of their $n$-bit inputs
is bigger. For this function, $\sR^{1, \pub}_\epsilon(\GT_n) =
\Theta(n)$ but $\vc(\GT_n) = 1$. Nonetheless for these functions, any
hard distribution $\mu$, is highly correlated between $\mcX$ and
$\mcY$. Therefore it is conceivable that such a relationship, as in
Eq.~\ref{eq:kremer}, could still hold, possibly after taking into
account the amount of correlation in a given non-product
distribution. This question, although probably never explicitly asked
in any previous work, appears to be quite fundamental. We answer it in
the positive by the following.
\begin{theorem}
\label{thm:vc-dim-non-product}
Let~$f : \mcX \times \mcY \rightarrow \{0,1\}$ be a boolean function
and let~$\epsilon \in (0, 1/2)$ be a constant. Let $\mu$ be a distribution (possibly
non-product) on $\mcX \times \mcY$. Let $XY$ be joint random
variables distributed according to $\mu$. There is a universal
constant~$\kappa$ such that,
\[ 
    \sD^{1,\mu}_{\epsilon}(f) 
    \quad \leq \quad \kappa \cdot \frac{1}{\epsilon}
               \log\frac{1}{\epsilon} \cdot
\left(\frac{1}{\epsilon} \cdot I(X:Y) + 1 \right) \cdot \vc(f) 
\]
In particular, for constant $\epsilon$,
\[  \sD^{1,\mu}_\epsilon(f) \quad =  \quad  O \left( (I(X:Y) + 1)
\cdot \vc(f) \right)\]  
Above $I(X:Y)$ represents the mutual information between correlated
random variables $X$ and $Y$, distributed according to $\mu$.  
\end{theorem}
Let us discuss below a few aspects of this result and its
relationship with what is previously known. Note that in combination
with  Yao's Principal, Thm.~\ref{thm:vc-dim-non-product} gives us
the following (where the mutual information is now considered under
a hard distribution for $f$).  
\begin{equation} 
\label{eq:randcc} 
\sR^{1,\pub}(f) = O \left((I(X:Y) + 1) \cdot \vc(f) \right). 
\end{equation}

\begin{enumerate} 
\item 
It is easily observed using Sauer's Lemma (Lem.~\ref{lem:sauer},
Sec.~\ref{sec:prelim}.) that the deterministic complexity of $f$ has
\begin{equation} \label{eq:detcc}  \sD^1(f) = O( \vc(f) \cdot \log
|\mcY|). \end{equation} 
This is because $\alice$ can simply tell the name of $f_x$ in $O(
\vc(f) \cdot \log |\mcY| )$ bits since $|\mcF| \leq |\mcY|^{\vc(f)}$.  Now
our result (\ref{eq:randcc}) is on one hand stronger than
(\ref{eq:detcc}) in the sense $I(X:Y) \leq
\log |\mcY|$ always, and $I(X:Y)$ could be much smaller than $\log |\mcY|$ depending
on $\mu$. An example of such a case is the Inner Product
($\mathsf{IP}_n$) function in which $\alice$ and $\bob$ need to
determine the inner product (mod $2$) of their $n$-bit input
strings. For $\mathsf{IP}_n$, a hard distribution is the uniform
distribution which is product, and hence 
$I(X:Y) = 0$, whereas $\log |\mcY| =n$. However on the other hand
(\ref{eq:randcc}) is also weaker than (\ref{eq:detcc}) in the sense it
only upper bounds the public-coin randomized complexity,
whereas (\ref{eq:randcc}) upper bounds the deterministic complexity of $f$. 

\item Aaronson~\cite{Aaronson07} shows that for a total or partial
boolean function $f$, 
\begin{equation} \label{eq:scott} \sR^1(f) =
O(\sQ^1(f) \cdot \log |\mcY|). \end{equation} Again (\ref{eq:randcc})
is stronger than (\ref{eq:scott}) in the sense that $I(X:Y)$ could be
much smaller than $\log |\mcY|$ depending on $\mu$. Also it is known
that, $\sQ^1(f) = \Omega( \vc(f))$ always, following from
Nayak~\cite{Nayak99}, and $\sQ^1(f)$ could be much larger than
$\vc(f)$. An example is the Greater Than ($\mathsf{GT}_n$) function
for which $\sQ^1(\mathsf{GT}_n) = \Omega(n)$, whereas
$\vc(\mathsf{GT}_n) = O(1)$. On the other hand (\ref{eq:randcc}) only
holds for total boolean functions whereas (\ref{eq:scott}) also holds
for partial boolean functions.

\item As mentioned before, for all total boolean functions $f$, $\sR^{1,\pub}(f) =
\Omega(\vc(f))$, and $\sR^{1,\pub}(f)$ could be much larger
than $\vc(f)$ (as in function $\mathsf{GT_n}$). 
Now Eq.~(\ref{eq:randcc}) says that in the latter case, the mutual information
$I(X:Y)$ under any hard distribution $\mu$ must be large. That is, a
hard distribution $\mu$ must be highly correlated. 

\item It is known that for total boolean functions $f$, for which a
hard distribution is product, there is no separation between the
one-way public-coin randomized and quantum communication
complexities. Now our theorem gives a  
 {\em smooth} extension of this fact to the functions whose hard
distributions are not product ones. 
\end{enumerate}

\vspace{0.1in}
\noindent A generalization of the $\vc$-dimension for non-boolean
functions, is referred to as the {\em pseudo-dimension}
(Def.~\ref{def:pseudo}, Sec.~\ref{sec:prelim}). For a
non-boolean function $f : \mcX
\times \mcY \rightarrow \set{1, \ldots, k}$ ($k \geq 2$ an integer), we show a
similar upper bound on $\sD_{\epsilon}^{1,\mu}(f)$ in terms of $k,
I(X:Y)$ and the pseudo-dimension of $f' \defeq \frac{f}{k}$. 
\begin{theorem} 
\label{thm:pdim-1}
Let $k\geq 2$ be an integer. Let~$f : \mcX \times \mcY \rightarrow
\set{1, \ldots, k}$ 
 and $\epsilon \in (0, 1/6)$ be a
constant. Let $f': \mcX \times \mcY \rightarrow [0,1]$ be such that $f'(x,y) =
f(x,y)/k$. Let $\mu$ be a distribution (possibly non-product) on $\mcX
\times \mcY$, and $XY$ be joint random variables distributed according
to $\mu$. Then there is a universal constant~$\kappa$ such that,
\[ 
    \sD^{1,\mu}_{3\epsilon}(f) \quad \leq \quad \kappa \cdot
    \frac{k^4}{\epsilon^5}\cdot \left( \log
\frac{1}{\epsilon} + d \log^2
\frac{dk}{\epsilon}\right) \cdot \left( I(X:Y) +  \log k \right)
\]
where $d \defeq \mcP_{\frac{\epsilon^2}{576k^2}}(f')$ is the
$\frac{\epsilon^2}{576k^2}$-pseudo-dimension of $f'$.  
\end{theorem}

\vspace{0.1in}
\noindent Let us now discuss our other main result which we show in
the quantum one-way model. 

\subsection{Quantum lower bound}
For a function $f: \mcX \times \mcY \rightarrow \mcZ$, a measure of
its complexity that is often very useful in understanding its
classical randomized communication complexity, is what is referred to
as the {\em rectangle bound} (denoted by $\rec(f)$), also often known
as the {\em corruption bound}.  The rectangle bound $\rec(f)$ is
actually defined first via a distributional version $\rec^\mu(f)$. It
is a well studied measure and $\rec^{\mu}(f)$ is well known to form a
lower bound on $\sD^\mu(f)$ both in the one-way and two-way models. In
fact, in a celebrated result, Razborov~\cite{Raz92} provided optimal
lower bound on the randomized communication complexity of the Set
Disjointness function, by arguing a lower bound on its rectangle
bound.

It is natural to ask if this measure also forms a lower bound on the
quantum communication complexity. We answer in the positive for this
question in the one-way model. We show that, for a total or partial
function, the quantum distributional one-way communication complexity
under a given product distribution $\mu$ is lower bounded by the
corresponding one-way rectangle bound. Our precise result is as
follows.

\begin{theorem} 
\label{thm:reclower-1}
Let $f : \mcX \times \mcY \rightarrow \mcZ$ be a
total function and let $\epsilon \in (0,1/2)$ be a constant.  Let $\mu$ be a
product distribution on $\mcX \times \mcY$ and let $
\rec_\epsilon^{1,\mu}(f) > 2\cdot \log
(1/\epsilon)$. Then, 
\begin{equation}
\sQ_{\epsilon^3/8}^{1, \mu}(f) \quad \geq \quad \frac{1}{2} \cdot (1 -
2\epsilon) \cdot (S(\epsilon/2) - 
S(\epsilon/4)) \cdot (\lfloor \rec_\epsilon^{1,\mu}(f) \rfloor - 1) =
\Omega(\rec_\epsilon^{1,\mu}(f)), 
\end{equation}
where for $p \in (0,1)$, $S(p)$ is the binary entropy function
$S(p) \defeq -p \log p - (1-p) \log (1-p)$.

If $f : \mcX \times \mcY \rightarrow \mcZ \cup \{*\}$ is a partial
function then,
$$ \sQ_{\epsilon^6/(2 \cdot 15^4)}^{1,
\mu}(f) \quad \geq \quad \frac{1}{2} \cdot (1 - 2\epsilon) \cdot
\frac{\epsilon^2}{300} \cdot (\lfloor \rec_\epsilon^{1,\mu}(f) \rfloor
- 1) = \Omega(\rec^{1,\mu}_\epsilon(f)).$$  
\end{theorem}

Let us make a few important remarks here related to this result.

\begin{enumerate}
\item Recently, Jain, Klauck and Nayak~\cite{JainKN08} showed that for any relation $f
\subseteq \mcX \times \mcY \times \mcZ$,
the rectangle bound of $f$ tightly characterizes the randomized one-way
classical communication complexity of $f$.
\begin{theorem} [\cite{JainKN08}]
\label{thm:Reqrecjkn} 
Let $f \subseteq \mcX \times \mcY \times \mcZ$ be a relation and let
$\epsilon \in (0,1/2)$. Then, $$\sR_\epsilon^{1, \pub}(f) =
\Theta(\rec_\epsilon^{1}(f)). $$
\end{theorem}

While showing Thm.~\ref{thm:Reqrecjkn},
Jain, Klauck and Nayak~\cite{JainKN08} have shown that for all relations
$f: \mcX \times \mcY \rightarrow \mcZ$ and for all distributions $\mu$
(product and non-product) on $\mcX \times \mcY$; $\sD_\epsilon^{1,
\mu}(f) = \Omega(\rec_{4\epsilon}^{1,\mu}(f)) $. However in the quantum
setting we are making a similar statement only for (total or partial)
functions $f$ and only for product
distributions $\mu$ on $\mcX \times \mcY$. In fact it {\bf does
NOT} hold if we let $\mu$ to be non-product. It can be shown
that there is a total function $f$ and a non-product distribution
$\mu$ such that $\sQ_\epsilon^{1,\mu}(f)$ is exponentially smaller than
$\rec_\epsilon^{1,\mu}(f)$. This fact is implicit in the work of Gavinsky
et al.~\cite{Gavinskyetal07}. We make an explicit statement of this in
Sec.~\ref{sec:example}. in Appendix and skip its proof for brevity.

\item Let $\epsilon \in (0,1/4)$. Jain, Klauck and
Nayak~\cite{JainKN08} have shown that for all relations $g \subseteq
\mcX \times \mcY \times \mcZ$, 
$$ \sR^{1,[]}_{2\epsilon}(g) \quad = \quad O(\rec^{1,
[]}_{\epsilon}(g)).$$ Here the superscript $[]$ represents
maximization over all product distributions. 
From Thm.~\ref{thm:reclower-1} for a (total or partial) function $f$ we
get, $$\sQ^{1,[]}_{\epsilon^6/(2 \cdot 15^4)}(f) \quad = \quad \Omega(\rec^{1,
[]}_\epsilon(f)).$$ 
Since $\sR^{1,[]}_\epsilon(f) \geq
\sQ^{1,[]}_\epsilon(f)$, combining everything we get,
\begin{theorem} Let $\epsilon \in (0,1/4)$. Let $f: \mcX \times
\mcY \rightarrow \mcZ \cup \set{*}$ be a (possibly partial and non-boolean)
function. Then $$\sR^{1,[]}_{\epsilon^6/(2 \cdot 15^4)}(f) \quad \geq \quad
\sQ^{1,[]}_{\epsilon^6/(2 \cdot 15^4)}(f) \quad = \quad \Omega(\sR^{1,
[]}_{2\epsilon}(f)).$$
\end{theorem} 
It was known earlier that for total boolean functions, $\sQ^{1,[]}(f)$
is tightly bounded by $\sR^{1,[]}(f)$. We extend such a relationship
here to apply for non-boolean (partial) functions as well. We remark that the
earlier proofs for total boolean functions used the $\vc$-dimension
result, Thm.~\ref{thm:vc-dim}, of Kremer, Nisan and
Ron~\cite{KremerNR95}. We get the same result here without requiring
it.
\end{enumerate}

\vspace{0.1in}
\noindent We finally present an application of our result
Thm.~\ref{thm:reclower-1} in the context of studying security of
extractors against {\em quantum adversaries}. An extractor is a
function that is used to extract almost uniform randomness from a
source of imperfect randomness. As very well studied
objects, extractors have found several uses in many cryptographic
applications and also in complexity theory. Recently, security of
various extractors has been increasingly studied in the presence of
quantum adversaries; since such secure extractors are then useful
in several applications such as privacy amplification in
quantum key distribution and key-expansion in quantum bounded storage
models~\cite{KonigMR05,KonigR05,KonigT08}. In particular, K{\"o}nig and
Terhal~\cite{KonigT08} have shown that any boolean extractor that can
extract a uniform bit from sources of {\em min-entropy} $k$ is also secure
against quantum adversaries with their memory bounded by a function of
$k$. 

We get a similar statement for boolean extractors, as a corollary of our result
Thm.~\ref{thm:reclower-1}. We obtain this corollary by observing a
key connection between the minimum min-entropy that an extractor function $f$
needs to extract a uniform bit and its rectangle bound. The precise
statement of our result, its relationship with the result
of~\cite{KonigT08}, and other detailed discussions are deferred to
Sec.~\ref{sec:qextract}. 

\subsection{Organization} 
In the following Sec.~\ref{sec:prelim} we discuss various
information theoretic preliminaries and the model
of one-way communication. In Sec.~\ref{sec:upperbound} we
present the upper bounds in the classical setting. In the following
Sec.~\ref{sec:qlowerbound} we present the lower bounds in the
quantum setting. The application concerning extractors is discussed in
Sec.~\ref{sec:qextract}. We finally
conclude with some open questions in Sec.~\ref{sec:conclude}.

\section{Preliminaries}
\label{sec:prelim}
\subsection{Information theory} 
In this section we present some information theoretic notations,
definitions and facts that we use in the rest of the paper. For an introduction
to classical and quantum information theory, we refer the reader to the
texts by Cover and Thomas~\cite{CoverT91} and Nielsen and
Chuang~\cite{NielsenC00} respectively. Most of the facts stated in this section
without proofs may be found in these books.

All logarithms in this paper are taken with base 2, unless otherwise
specified. For an integer $t
\geq 1$, $[t]$ represents the set $\{1, \ldots, t\} $. For square
matrices $P,Q$, by $Q \geq P $ we mean that $Q - P $ is positive
semi-definite.  For a matrix $A$, $\norm{A}_1 \defeq \Tr
(\sqrt{A^\dagger A})$ denotes its $\ell_1$ norm. For $p \in (0,1)$,
let $S(p) \defeq -p \log p - (1-p) \log (1-p)$, denote the binary
entropy function.  We have the following fact.
\begin{fact}
\label{fact:ssmall}
For $\delta \in [0, 1/2], \quad S(\frac{1}{2} + \delta) \leq 1 - 2\delta^2$
and $S(\delta) \leq 2\sqrt{\delta}$.
\end{fact}

A quantum state, usually represented by letters $\rho, \sigma$ etc.,
is a positive semi-definite trace one operator in a given Hilbert
space.  Specializing from the quantum case, we view a discrete
probability distribution $P$ as a positive semi-definite trace one
diagonal matrix indexed by its (finite) sample space. For a
distribution $P$ with support on set $\mcX$, and~$x \in \mcX$, $P(x)$
denotes the $(x,x)$ diagonal entry of $P$, and~$P(\mcE) \defeq \sum_{x \in
\mcE} P(x)$ denotes the probability of the event~$\mcE \subseteq
\mcX$. A distribution $P$ on $\mcX \times \mcY$ is said to be {\em product}
across $\mcX$ and $\mcY$, if it can be written as $P = P_{\mcX}
\otimes P_{\mcY}$, where $P_\mcX, P_\mcY$ are distributions on $\mcX,
\mcY$ respectively and $\otimes$ is the tensor operation. Often for
product distributions we do not mention 
the sets across which it is product if it is clear from the context.

Let $X$ be a classical random variable (or simply random variable)
taking values in $\mcX$. For a random variable $X$, we also let $X$
represent its probability distribution. The {\em entropy} of $X$
denoted $S(X)$ is defined to be $S(X) \defeq - \Tr X
\log X$. Since $X$ is classical an equivalent definition would be
$S(X) \defeq - \sum_{x\in \mcX} \Pr[X = x] 
\log \Pr[X=x]$ . Let $X, Y$ be a correlated random variables taking
values in $\mcX, \mcY$ respectively. $XY$ are
said to be {\em independent} if their joint distribution is
product. The {\em mutual information} between them, denoted $I(X:Y)$
is defined to be $I(X:Y) \defeq S(X) + S(Y) - S(XY)$ and {\em
conditional entropy} denoted $S(X \vert Y)$ is defined to be $S(X
\vert Y) \defeq S(XY) - S(Y)$. It is easily seen that $S(X \vert Y) =
\av_{y \leftarrow Y}[S(X \vert (Y=y)]$.  

We have the following facts.

\begin{fact}
\label{fact:inf}
For all random variables $X,Y; ~I(X:Y) \geq 0$; in other words
$S(X) + S(Y) \geq S(XY)$. If $X,Y$ are independent then we have
$I(X:Y)=0$; in other words $S(XY) = S(X) + S(Y)$. 
\end{fact}

The definitions and facts stated in the above paragraph for classical random
variables also hold {\em mutatis mutandis} for quantum states as
well. For example for a quantum state $\rho$, its entropy is defined
as $S(\rho) \defeq -\Tr \rho \log \rho$. For
brevity, we avoid making all the corresponding statements
explicitly. As is the case with
classical random variables, for a quantum system say $Q$, we also
often let $Q$ represent its quantum state.  We have the following
fact.
\begin{fact}
\label{fact:qinf}
Any quantum state $\rho$ in $m$-qubits has $S(\rho) \leq m$. Also let $XQ$
be a joint classical-quantum system with $X$ being a classical random
variable, then
$I(X:Q) \leq \min \{S(X), S(Q)\}$.
\end{fact}

For a system $XYM$, let us define
$I(X:M \vert Y) \defeq S(X \vert Y) + S(M \vert Y) - S(XM \vert Y)$.
If $Y$ is a classical system then it is 
easily seen that $I(X:M \vert Y) = \av_{y \leftarrow Y} [I(X:M \vert
(Y=y))]$.

For random variables $X_1, \ldots, X_n$ and a correlated (possibly quantum) system
 $M$, we have the
following {\em chain rule of mutual information}, which will be
crucially used in our proofs.
\begin{equation} 
\label{eqn:chain}
I(X_1 \ldots X_n : M ) \quad = \quad \sum_{i= 1}^n
I(X_i : M | X_1 \ldots X_{i-1}) 
\end{equation}
By convention, conditioning on $X_1
\ldots X_{i-1}$ for $i=1$ means conditioning on the true event.

The following is an important information theoretic fact known as
Fano's inequality, which relates the probability of disagreement for
correlated random variables to their mutual information.

\begin{lemma}[Fano's inequality]
\label{lem:fano}
Let $X$ be a random variable taking values in $\mcX$. Let $Y$ be a
correlated random variable and let $P_e \defeq \Pr(X \neq Y)$. Then,
\begin{displaymath}
S(P_e) + P_e \log(\vert \mcX \vert - 1) \quad \geq \quad S(X\vert Y). 
\end{displaymath}
\end{lemma}

The VC-dimension of a boolean function $f$ is an important
combinatorial concept and has close connections with the one-way
communication complexity of $f$.
\begin{definition}[Vapnik-Chervonenkis ($\mathsf{VC}$) dimension] 
A set $S \subseteq \mcY$ is said to be {\em shattered} by a set $\mcG$
of boolean functions from $\mcY$ to $\{0,1\}$, if $\forall R \subseteq
S, \exists g_R \in \mcG$ such that $\forall s \in S, (s \in R)
\Leftrightarrow (g_R(s) = 1)$. The largest value $d$ for which there
is a set $S$ of size $d$ that is shattered by $\mcG$ is the {\em
Vapnik-Chervonenkis dimension} of $\mcG$ and is denoted by
$\vc(\mcG)$.

Let $f : \mcX \times \mcY \rightarrow \{0,1\}$ be a boolean
function. For all $x \in \mcX$ let $f_x: \mcY \rightarrow \{0,1\}$ be
defined as $ f_x(y) \defeq f(x,y), \forall y \in \mcY$. Let $\mcF \defeq
\{f_x : x \in \mcX \}$. Then the Vapnik-Chervonenkis dimension of $f$,
denoted by $\vc(f)$, is defined to be $\vc(\mcF)$.
\end{definition}

Let $f$ and $\mcF$ be as defined in the above definition. We call a
function $f$ {\em trivial} iff $|\mcF| = 1$, in other words iff the
value of the function, for all $x$, is determined only by $y$. We call
$f$ {\em non-trivial} iff it is not trivial. Note that a boolean $f$ is
non-trivial if and only if $\vc(f) \geq 1$.  Throughout this paper we
assume all our functions to be non-trivial.

Following is a useful fact, with several applications, relating
the VC-dimension of $f$ to the size of $\mcF$. It is usually
attributed to Sauer~\cite{Sauer72}, however it has been independently
discovered by several different people as well.
\begin{lemma}[Sauer's Lemma~\cite{Sauer72}]
\label{lem:sauer}
Let $f : \mcX \times \mcY \rightarrow \{0,1\}$ be a boolean
function. Let $d \defeq \vc(f)$. Let $m \defeq |\mcY|$, then
$$|\mcF| \quad \leq \quad \sum_{i=0}^d {m \choose i} \quad \leq \quad m^d. $$
\end{lemma}

The following result from Blumer, Ehrenfeucht, Haussler, and
Warmuth~\cite{BlumerEHW89} is one of the most fundamental results
from computational learning theory and in fact an important
application of Sauer's Lemma.

\begin{lemma}
\label{lem:learning}
  Let $H$ be class of boolean functions over a finite domain $\mcY$ with
VC-dimension $d$, let $\pi$ be an arbitrary probability distribution
over $\mcY$, and let $0 < \epsilon, \delta < 1$. Let L be any
algorithm that takes as input a set $S \in \mcY^m$ of $m$ examples
labeled according to an unknown function $h \in H$, and outputs a
hypothesis function $h'\in H$ that is consistent with $h$ on the
sample $S$. If $L$ receives a random sample of size $m \geq
m_0(d,\epsilon,\delta)$ distributed according to $\pi^m$, where 
  \begin{equation*}
    m_0(d, \epsilon, \delta) = c_0\left(\frac{1}{\epsilon} \log
\frac{1}{\delta} + \frac{d}{\epsilon}  \log \frac{1}{\epsilon}\right) 
  \end{equation*}
  for some constant $c_0 > 0$, then with probability at least $1 -
\delta$ over the random samples, $\Pr_\pi[h'(y) \neq h(y)] \leq
\epsilon$. 
\end{lemma}

A similar learning result also holds for non-boolean functions. For
this let us first define the following generalization of the
$\vc$-dimension, known as the {\em pseudo-dimension}.

\begin{definition}[pseudo-dimension]
\label{def:pseudo}
A set $S \subseteq \mcY$ is said to be
{\em $\gamma$-shattered} by a set $\mcG$ of functions from $\mcY$ to $\mcZ
\subseteq \rea$, if there exists a vector $\bw = (w_1, \ldots, w_k)
\in \mcZ^k$ of dimension $k = |S|$ for which the following holds. 
For all $ R \subseteq S, \exists g_R \in \mcG$ such
that $\forall s \in S, (s \in R) \Rightarrow (g_R(s) > w_i + \gamma)$
and $(s \notin R) \Rightarrow (g_R(s) < w_i - \gamma)$. The largest
value $d$ for which there is a set $S$ of size $d$ that is
$\gamma$-shattered by $\mcG$ is the {\em $\gamma$-pseudo-dimension} of
$\mcG$ and is denoted by $\pg(\mcG)$. 

Let $f: \mcX \times \mcY \rightarrow \mcZ$ be a function. For all $x
\in \mcX$ let $f_x: \mcY \rightarrow \mcZ$ be defined as $ f_x(y)
\defeq f(x,y), \forall y \in \mcY$. Let $\mcF \defeq
\{f_x : x \in \mcX \}$. Then the $\gamma$-pseudo-dimension of $f$,
denoted by $\pg(f)$, is defined to be $\pg(\mcF)$.
\end{definition}

Following result of Bartlett, Long and Williamson~\cite{BartlettLW96} is similar to the
learning lemma of Blumer et al.~\cite{BlumerEHW89} and concerns
non-boolean functions.

\begin{theorem}
\label{thm:learnnonbool}
Let $\mcG$ be a class of functions over a finite domain $\mcY$ into the range
$[0,1]$. Let $\pi$ be an arbitrary probability distribution over
$\mcY$ and let $\epsilon \in (0,1/2)$ and $\delta \in (0,1)$. Let $d
\defeq \mcP_{\epsilon^2/576}(\mcG)$. Then there exists a deterministic
learning algorithm $L$ which has the following property. Given as
input a set $S \in \mcY^m$ of $m$ examples chosen according to
$\pi^m$ and labeled according to an unknown function $g \in \mcG$, $L$
outputs a hypothesis $g' \in \mcG$ such that if $m \geq m_0(d,
\epsilon, \delta)$ where
$$ m_0(d,\epsilon, \delta) \quad = \quad c_0\left( \frac{1}{\epsilon^4} \log
\frac{1}{\delta} + \frac{d}{\epsilon^4}\log^2 \frac{d}{\epsilon}\right)$$
for some constant $c_0 > 0$, then with probability at least $1 -
\delta$ over the random samples,
$$ \sum_{y \in \mcY} \pi(y) \cdot |h'(y) - h(y)|  \quad \leq \quad \epsilon.$$ 
\end{theorem}

Following is a very fundamental  quantum information
theoretic fact shown by Holevo~\cite{Holevo73}.
\begin{theorem}[The Holevo bound~\cite{Holevo73}]
\label{thm:Holevo} 
Let $X$ be classical random variable taking values in $\mcX$. Let $M$
be a correlated quantum system and let $Y$ be a random variable
obtained by performing a quantum measurement on $M$. Then,
\begin{equation}
    I(X:Y) \quad \leq \quad I(X:M). 
\end{equation}
\end{theorem}

Following is an  interesting and useful information theoretic fact first
shown by Helstrom~\cite{Helstrom76}.

\begin{theorem}[\cite{Helstrom76}]
\label{thm:helstrom}
Let $XQ$ be joint classical-quantum system where $X$ is a classical
boolean random variable. For $a \in \{0,1\}$, let the quantum state of $Q$ when $X=a$ be
$\rho_a$. The optimal success probability of predicting $X$ with a
measurement on $Q$ is given by 
$$ \frac{1}{2} + \frac{1}{2} \cdot \|\Pr[X=0] \rho_0 - \Pr[X=1] \rho_1
\|_1.$$  
\end{theorem}

\subsection{One-way communication} 
\label{sec:commprelim}
In this article we only consider the two-party one-way model of
communication. Let $f \subseteq \mcX \times \mcY \times \mcZ$ be a
relation. The relations we consider are always total in the sense that for
every $(x,y) \in \mcX \times \mcY$, there is at least one $z \in
\mcZ$, such that $(x,y,z) \in f$. In a one-way protocol $\mcP$ for
computing $f$, $\alice$ and $\bob$ get inputs $x \in \mcX$ and $y \in
\mcY$ respectively. $\alice$ sends a single message to $\bob$, and
their intention is to determine an answer $z \in \mcZ$ such that
$(x,y,z) \in f$.  In the one-way protocols we consider, the single
message is always from $\alice$ to $\bob$. A total function $f: \mcX \times
\mcY \rightarrow \mcZ$, can be viewed as a special type of relations
in which for every $(x,y)$ there is a unique $z$, such that $(x,y,z)
\in f$. A partial function is a special type of relations such that
for some inputs $(x,y)$, there is a unique $z$, such that $(x,y,z)
\in f$ and for all other inputs  $(x,y)$, $ (x,y,z)
\in f, \forall z \in \mcZ$. We view a partial function $f$ as a function 
$f: \mcX \times \mcY \rightarrow \mcZ \cup \set {*}$, such that the
inputs $(x,y)$ for which $f(x,y)= *$ are exactly the ones for which $ (x,y,z)
\in f, \forall z \in \mcZ$. 
 
Let us first consider classical communication protocols.  We let
$\sD^1(f)$ represent the deterministic one-way communication
complexity, that is the communication of the best deterministic
protocol computing $f$ correctly on all inputs. For~$\epsilon \in
(0,1/2)$, let~$\mu$ be a probability distribution on~$\mcX
\times \mcY$. We let $\sD_{\epsilon}^{1,\mu}(f)$ represent the 
distributional one-way communication complexity of $f$ under $\mu$
with expected error $\epsilon$, i.e., the communication of the best
private-coin one-way protocol for $f$, with distributional
error (average error over the coins and the inputs) at most
$\epsilon$ under $\mu$.  It is easily noted that $\sD_{\epsilon}^{1,\mu}(f)$ is
always achieved by a deterministic one-way protocol, and will henceforth
restrict ourselves to deterministic protocols in the context of
distributional communication complexity.  We let
$\sR^{1,\pub}_{\epsilon}(f)$ represent the public-coin randomized
one-way communication complexity of $f$ with worst case error
$\epsilon$, i.e., the communication of the best public-coin randomized
one-way protocol for $f$ with error for each input $(x,y)$ being at
most~$\epsilon$.  The analogous quantity for private coin randomized
protocols is denoted by $\sR^{1}_{\epsilon}(f)$. The public- and private-coin randomized communication complexities are not much different, as shown in Newman's result~\cite{New91} that 
\begin{equation}
	\sR^1(f) = O(\sR^{1,pub}(f)+ \log\log |\mcX| + \log \log |\mcY|).
\end{equation} 
The following result
due to Yao~\cite{Yao77} is a very useful fact connecting worst-case and distributional communication complexities. It is a
consequence of the {\em min-max\/} theorem in game
theory~\cite[Thm.~3.20, page~36]{KushilevitzN97}.
\begin{lemma}[Yao's principle~\cite{Yao77}]
\label{lem:yao} $\sR^{1,\pub}_{\epsilon}(f) = \max_{\mu}
\sD_{\epsilon}^{1,\mu}(f)$.
\end{lemma}

We define~$\sR^{1,[]}_{\epsilon}(f) \defeq \max_{\mu \textrm{ product}}
\sD_{\epsilon}^{1,\mu}(f)$. Note that $\sR^{1,[]}_{\epsilon}(f)$ could
be significantly smaller than $\sR^{1, \pub}_{\epsilon}(f)$ as is
exhibited by the Greater Than ($\mathsf{GT_n}$) function for which
$\sR^{1,\pub}(\mathsf{GT}_n) = \Omega(n)$, whereas
$\sR^{1,[]}_{\epsilon}(f) = O(1)$.

In a one-way quantum communication protocol, $\alice$ and $\bob$ are
allowed to do quantum operations and $\alice$ can send a quantum
message (qubits) to $\bob$. Given $\epsilon \in (0,1/2)$, the one-way
quantum communication complexity $\sQ^1_\epsilon(f)$ is defined to be
the communication of the best one-way quantum protocol with error at most
$\epsilon$ on all inputs. Given a distribution $\mu$ on $\mcX \times
\mcY$, we can similarly define the quantum  distributional one-way
communication complexity of $f$, denoted $\sQ^{1,\mu}_\epsilon(f)$, to be  
the communication of the best one-way quantum protocol $\mcP$ for $f$ such
that the average error of $\mcP$ over the inputs drawn from the
distribution $\mu$ is at most $\epsilon$. We
define~$\sQ^{1,[]}_{\epsilon}(f) \defeq \max_{\mu \textrm{ product}} 
\sQ_{\epsilon}^{1,\mu}(f)$.

\section{A new upper bound on classical  one-way distributional
communication complexity} 
\label{sec:upperbound}
In this section we present the upper bounds
on the distributional communication complexity,
$\sD^{1,\mu}_\epsilon(f)$ for any distribution $\mu$ (possibly
non-product) on $\mcX \times \mcY$. We begin by restating the precise
result for boolean functions.

\begin{theorem} 
\label{thm:vc-dim-non-product-1}
Let~$f : \mcX \times \mcY \rightarrow \{0,1\}$ be a boolean function
and let~$\epsilon \in (0, 1/2)$ be a constant. Let $\mu$ be a distribution (possibly
non-product) on $\mcX \times \mcY$. Let $XY$ be joint random
variables distributed according to $\mu$. There is a universal
constant~$\kappa$ such that,
\[ 
    \sD^{1,\mu}_{\epsilon}(f) 
    \quad \leq \quad \kappa \cdot \frac{1}{\epsilon}
               \log\frac{1}{\epsilon} \cdot
\left(\frac{1}{\epsilon}\cdot I(X:Y) + 1\right) \cdot \vc(f). 
\]
In other words,
\[  \sD^{1,\mu}_{\epsilon}(f) 
    \quad =  \quad  O \left( (I(X:Y) + 1) \cdot \vc(f) \right)\] 
\end{theorem}

For showing this result we will crucially use the following fact shown
by Harsha, Jain, McAllester and Radhakrishnan~\cite{HarshaJMR07}
concerning communication required for generating correlations. We
begin with the following definition.

\begin{definition}[Correlation protocol]
\label{def:corr}
Let $(X,Y)$ be a pair of correlated random variables taking values in
$\mathcal X \times \mathcal Y$. Let $\alice$ be given $x \in \mathcal X$,
sampled according to the distribution X. $\alice$ should transmit a
message to $\bob$, such that $\alice$ and $\bob$ can together generate a value
$y\in \mathcal Y$ distributed according to the conditional
distribution $Y\vert_{X=x}$; that is the pair $(x, y)$ should have joint
distribution $(X,Y)$. $\alice$ and $\bob$ are allowed to use public
randomness. Note that the generated value $y$ should be known to both $\alice$
and $\bob$. 
\end{definition}

Harsha et al.~\cite{HarshaJMR07} showed that the minimal expected number of
bits that $\alice$ needs to send (in the presence of shared randomness),
denoted $T^R(X:Y)$, is characterized by the mutual information
$I(X:Y)$ as follows.
\begin{theorem}[\cite{HarshaJMR07}]
\label{thm:corr}
There exists a universal positive constant $l$ such that, 
$$ I(X:Y) \quad \leq \quad  T^R(X:Y) \quad \leq \quad 
4I(X :  Y) + l.  $$
\end{theorem}

We will also need the following fact. 
\begin{lemma}
\label{lem:infadd}
Let $m \geq 1$ be an integer. Let $XY$ be correlated random
variables. Let $\mu_x$ be the distribution of $Y \vert X=x$. Let
$X'Y'$ represent joint random variables such that $X'$ is distributed
identically to $X$ and the distribution of $Y'\vert (X'=x)$ is
$\mu_x^{\otimes m}$ ($m$ independent copies of $\mu_x$).  Then, $$
I(X':Y') \quad \leq \quad m \cdot I(X:Y).$$
\end{lemma}

\begin{proof}
Consider,
\begin{eqnarray*}
I(X':Y') & = &  S(Y') - \av_{x \leftarrow X'} [S(Y' \vert X' =x)] \\
& = & S(Y') - m \cdot \av_{x \leftarrow X} [S(Y \vert X =x)] \\
& \leq & m \cdot S(Y) - m \cdot  \av_{x \leftarrow X} [S(Y \vert X
=x)] \\
& = & m \cdot I(X:Y)
\end{eqnarray*}
The second equality above  follows from
Fact~\ref{fact:inf} and since $X'$ and $X$ are identically 
distributed. Similarly the first inequality above follows from
Fact~\ref{fact:inf} by noting that $Y'$ is  $m$-copies of $Y$.
\end{proof}

We are now ready for the proof of Thm.~\ref{thm:vc-dim-non-product-1}.

\begin{proofof}{Thm.~\ref{thm:vc-dim-non-product-1}}
Let $m \defeq  m_0(\vc(f), \epsilon/4, \epsilon/4) =  c_0 \cdot
\left(\frac{1}{\epsilon/4} \log \frac{1}{\epsilon/4} \right) \cdot (
\vc(f) + 1)$ as in  Lem.~\ref{lem:learning}. Let $l$ be the constant as in
Thm.~\ref{thm:corr}. Let $c \defeq 4m \cdot I(X:Y) + l$. We exhibit a
public coin protocol $\mcP$ with inputs drawn from $\mu$, in which
$\alice$ sends two messages $M_1$ and $M_2$ to $\bob$. The expected
length of $M_1$ is at most $c$ and the length of $M_2$ is always at
most $m$. The average error (over inputs and coins) of $\mcP$ is at
most $\epsilon/2$. Let $\mcP'$ be the protocol that simulates $\mcP$
but aborts and outputs $0$, whenever the length of $M_1$ in $\mcP$
exceeds $2c/\epsilon$. From Markov's inequality this happens with
probability at most $\epsilon/2$. Hence the expected error of $\mcP'$
is at most $\epsilon/2 + \epsilon/2 =
\epsilon.$ From $\mcP'$, we finally get a 
deterministic protocol with communication bounded by $2c/\epsilon + m$ and
distributional error at most $\epsilon$.  This implies our result from
definition of $\sD^{1,\mu}_{\epsilon}(f)$ and by setting $\kappa$ appropriately. 

 For $x \in \mcX$, let $\mu_x$ be the
distribution of $Y \vert X=x$. In $\mcP$, on receiving the input $x
\in \mcX$, $\alice$ first sends  a message $M_1$ to $\bob$, according
to the corresponding correlation protocol as in
Definition~\ref{def:corr}, and they together sample from the
distribution of $\mu_x^{\otimes m}$. Let $y_1, \ldots , y_m$ be the
samples generated. Note that from the properties of correlation
protocol both $\alice$ and $\bob$ know the values of $y_1, \ldots ,
y_m$. $\alice$ then sends to $\bob$ the second message $M_2$ which is
the values of $f(x,y_1),
\ldots , f(x,y_m)$. $\bob$ then considers the first  $x'$ (according
to the increasing order) such that $\forall
i \in [m], f(x',y_i) = f(x,y_i)$ and outputs $f(x',y)$, where $y$ is
his actual input. Using Lem.~\ref{lem:learning}, it is easy to verify
that for every $x \in \mcX$, the average error (over randomness in the
protocol and inputs of $\bob$) in this protocol $\mcP$ will be at most
$\epsilon/2$. Hence also the overall average error of $\mcP$ is at
most $\epsilon/2$. Also from Thm.~\ref{thm:corr} and
Lem.~\ref{lem:infadd}, we can verify that the expected length of
$M_1$ in $\mcP$ will be at most $4m \cdot I(X:Y)+ l$.  
\end{proofof}

\vspace{0.1in}
Following similar arguments and using Thm.~\ref{thm:learnnonbool} and
Thm.~\ref{thm:corr}, we obtain a similar result for
non-boolean functions as follows.

\begin{theorem} 
\label{thm:pdim}
Let $k\geq 2$ be an integer. Let~$f : \mcX \times \mcY \rightarrow
[k]$ be a non-boolean function and let~$\epsilon \in (0, 1/6)$ be a
constant. Let $f': \mcX \times \mcY \rightarrow [0,1]$ be such that $f'(x,y) =
f(x,y)/k$. Let $\mu$ be a distribution (possibly non-product) on $\mcX
\times \mcY$. Let $XY$ be joint random variables distributed according
to $\mu$. There is a universal constant~$\kappa$ such that,
\[ 
    \sD^{1,\mu}_{3\epsilon}(f) \quad \leq \quad \kappa \cdot
    \frac{k^4}{\epsilon^5}\cdot \left( \log
\frac{1}{\epsilon} + d \log^2
\frac{dk}{\epsilon}\right) \cdot \left( I(X:Y) +  \log k \right)
\]
where $d \defeq \mcP_{\frac{\epsilon^2}{576k^2}}(f')$ is the
$\frac{\epsilon^2}{576k^2}$-pseudo-dimension of $f'$.   
\end{theorem}

\begin{proof}
Let $m \defeq  m_0(d, \epsilon/k, \epsilon) =  c_0\left( \frac{k^4}{\epsilon^4} \log
\frac{1}{\epsilon} + \frac{dk^4}{\epsilon^4}\log^2
\frac{dk}{\epsilon}\right)$ as in  Thm.~\ref{thm:learnnonbool}. Let
$l$ be the constant as in Thm.~\ref{thm:corr}. Let $c \defeq 4m
\cdot I(X:Y) + l$. We exhibit a public coin protocol $\mcP$ for $f$,
with inputs drawn from $\mu$, in which $\alice$ sends two messages
$M_1$ and $M_2$ to $\bob$. The expected length of $M_1$ is at most $c$
and the length of $M_2$ is always at most $O(m \log k )$. The
average error (over inputs and coins) of $\mcP$ is at most
$2\epsilon$. Let $\mcP'$ be the protocol that simulates $\mcP$ but
aborts and outputs $0$, whenever the length of $M_1$ in $\mcP$ exceeds
$c/\epsilon$. From Markov's inequality this happens with probability
at most $\epsilon$. Hence the expected error of $\mcP'$ is at most
$2\epsilon + \epsilon = 3\epsilon.$ From $\mcP'$, we finally get a
deterministic protocol with communication bounded by $c/\epsilon + O(m
\log k)$
and distributional error at most $3\epsilon$.  This implies our result
from definition of $\sD^{1,\mu}_{3\epsilon}(f)$ and by setting
$\kappa$ appropriately.

In $\mcP$, $\alice$ and $\bob$ intend to first determine $f'(x,y)$ and
then output $kf'(x,y)$.  For $x \in \mcX$, let $\mu_x$ be the
distribution of $Y \vert X=x$.  On receiving the input $x
\in \mcX$, $\alice$ first sends  a message $M_1$ to $\bob$, according
to the corresponding correlation protocol as in
Definition~\ref{def:corr}, and they together sample from the
distribution of $\mu_x^{\otimes m}$. Let $y_1, \ldots , y_m$ be the
samples generated.  $\alice$ then sends to $\bob$ the second message
$M_2$ which is the values of $f'(x,y_1),
\ldots , f'(x,y_m)$ . $\bob$ then considers $x'$ as obtained from the
learning algorithm $L$ (as in Thm.~\ref{thm:learnnonbool}) and then
outputs $kf'(x',y)$, where $y$ is his actual input. Therefore from
Thm.~\ref{thm:learnnonbool}, with probability $1 - \epsilon$ over the
samples $y_1, \ldots , y_m$,
\begin{equation}
\label{eq:smalldiff}
\sum_{y \in
\mcY} \pi(y) \cdot |f'(x',y) - f'(x,y)| \quad \leq \quad  \epsilon/k.
\end{equation}
Note that, $(f'(x',y) \neq f'(x,y)) \Rightarrow |f'(x',y) - f'(x,y)|
\geq 1/k$. Hence for samples $y_1, \ldots , y_m$, for which
(\ref{eq:smalldiff}) holds, using Markov's inequality, we have $\Pr_{y
\leftarrow \mu_x}[f'(x',y) \neq f'(x,y)] \leq \epsilon$.
Therefore, for any fixed $x$, the error of $\mcP$ is at most $2\epsilon$
and hence also the overall error of $\mcP$ is at most $2\epsilon$.

From Thm.~\ref{thm:corr} and Lem.~\ref{lem:infadd}, we can
verify that the expected length of $M_1$ in $\mcP$ will be at most $4m
\cdot I(X:Y)+ l$. The length of $M_2$ is at most $O(m \log k)$, since
using a prefix free encoding each $f'(x,y_i)$ can be specified in $O(\log
k)$ bits.
  
\end{proof}

\section{A new lower bound on quantum one-way  distributional communication complexity}

\label{sec:qlowerbound}

In this section we present our lower bound on the quantum one-way
distributional communication complexity of a function $f$, in terms of
the one-way rectangle bound of $f$. We begin with a few definitions
leading to the definition of the one-way rectangle bound.
\begin{definition}[Rectangle]
A {\em one-way rectangle} $R$ is a set $S \times \mcY$, where $S \subseteq
\mcX$. For a distribution $\mu$ over $\mathcal{X} \times
\mathcal{Y}$, let $\mu_R$  represent the distribution arising from
$\mu$ conditioned on the event $R$ and let $\mu(R)$ represent the probability
(under $\mu$) of
the event $R$.
\end{definition}

\begin{definition}[One-way $\epsilon$-monochromatic] 
\label{def:mono}
Let $f \subseteq \mcX \times \mcY \times \mcZ$ be a relation. We call a
distribution $\lambda$ on $\mcX \times \mcY$, {\em one-way
$\epsilon$-monochromatic\/} for $f$ if there is a function~$g : \mcY
\rightarrow \mcZ$ such that $\Pr_{XY \sim \lambda} [ (X,Y,g(Y)) \in f ]
\geq 1 - \epsilon$.
\end{definition}

\begin{definition}[Rectangle bound]
\label{def:rec}
Let $f \subseteq \mcX \times \mcY \times \mcZ$ be a relation. For
distribution $\mu$ on $\mathcal{X} \times \mathcal{Y}$, the {\em one-way
rectangle bound} is defined as: $$ \rec_\epsilon^{1,\mu}(f) \defeq \min
\{ \log_2 \frac{1}{\mu(R)}: R
\text{ is one-way rectangle and $\mu_R$ is one-way
$\epsilon$-monochromatic} \}. $$
The {\em one-way rectangle bound} for $f$ is defined as: 
$$ \rec^1_\epsilon(f) \defeq \max_\mu  \rec_\epsilon^{1,\mu}(f).$$
We also define,
$$ \rec^{1, []}_\epsilon(f) \defeq \max_{\mu: \mathrm{product}}
\rec_\epsilon^{1,\mu}(f).$$ 
\end{definition}

We restate our precise result here followed by its proof. 
\begin{theorem} 
\label{thm:reclower}
Let $f : \mcX \times \mcY \rightarrow \mcZ$ be a
total function and let $\epsilon \in (0,1/2)$ be a constant.  Let $\mu$ be a
product distribution on $\mcX \times \mcY$ and let $ \rec_\epsilon^{1,\mu}(f) > 2(\log
(1/\epsilon))$. Then, $$ \sQ_{\epsilon^3/8}^{1,
\mu}(f) \quad \geq \quad \frac{1}{2} \cdot (1 - 2\epsilon) \cdot (S(\epsilon/2) -
S(\epsilon/4)) \cdot (\lfloor \rec_\epsilon^{1,\mu}(f) \rfloor - 1).$$ 
If $f : \mcX \times \mcY \rightarrow \mcZ \cup \{*\}$ is a partial
function then,
$$ \sQ_{\epsilon^6/(2 \cdot 15^4)}^{1,
\mu}(f) \quad \geq \quad \frac{1}{2} \cdot (1 - 2\epsilon) \cdot
\frac{\epsilon^2}{300} \cdot (\lfloor \rec_\epsilon^{1,\mu}(f) \rfloor - 1).$$  
\end{theorem}

We begin with the following information theoretic fact.
\begin{lemma}
\label{lem:largeinf}
Let $0 \leq d < c \leq 1/2$. Let $Z$ be a binary random variable
with~$\min\{\Pr(Z=0), \Pr(Z=1)\} \geq c$. Let $M$ be a correlated quantum
system. Let $Z'$ be a classical boolean random variable obtained by performing a
measurement on $M$ such that, $\Pr(Z \neq Z') \leq d$, then 
$$I(Z:M) \quad \geq \quad I(Z:Z') \quad \geq \quad S(c) - S(d).$$ 
\end{lemma}

\begin{proof}
The first inequality follows from the Holevo bound,
Thm.~\ref{thm:Holevo}. For the second inequality we note that $S(Z)
\geq S(c)$ (since the binary entropy function is monotonically
increasing in $(0,1/2]$) and from Fano's inequality, Lem.~\ref{lem:fano}, we have
$S(Z|Z') \leq S(d)$. Therefore, 
$$ I(Z:Z') \; = \; S(Z) - S(Z \vert Z') \; \geq \; S(c) - S(d). $$
\end{proof}

We are now ready for the proof of Thm.~\ref{thm:reclower}.
\\ \\
\begin{proofof}{Thm.~\ref{thm:reclower}}

\noindent {\bf For total boolean functions:}
 For simplicity of the explanation, we first present the proof assuming
$f$ to be a total boolean function. Let $r \defeq
\lfloor \rec_\epsilon^{1,\mu}(f) \rfloor $ or $\lfloor
\rec_\epsilon^{1,\mu}(f) \rfloor - 1$ so as to make $r$ even. Let $\mcP$ be the optimal
one-way quantum protocol 
for $f$ with distributional error under $\mu$ at most
$\epsilon^3/4$. (Although we have made a stronger assumption regarding
the error in the statement of the Theorem, we do not need it here and
will only need it  later while handling non-boolean functions.) Let
$M$ represent the $m \defeq \sQ_{\epsilon^3/4}^{1, \mu}(f)$ qubit
quantum message of $\alice$ in $\mcP$.  Let $XY$ be the random
variables corresponding to $\alice$ and $\bob$'s inputs, jointly
distributed according to $\mu$.  Our intention is to define binary
random variables $T_1,
\ldots, T_{r/2}$ such that they are determined by $X$ (and hence a
specific value for $T_1, \ldots, T_{r/2}$ would correspond to a subset
of $\mcX$) and $\forall i \in \{0, \ldots, \frac{r}{2}-1 \}$, $$ I(M :
T_{i+1} \vert T_1 \ldots T_i) \quad \geq \quad (1 - 2\epsilon) \cdot
(S(\epsilon/2) - S(\epsilon/4)).$$ 
Therefore from Fact~\ref{fact:qinf} and the chain rule of mutual information,
Eq.~(\ref{eqn:chain}), we have,
\begin{eqnarray*}
m   \geq S(M)  & \geq & I(M: T_1 \ldots T_{r/2}) \\ 
& = & \sum_{i=0}^{r/2-1} I(M : T_{i+1} \vert T_1 \ldots T_i) \\
& \geq &  (1 - 2\epsilon) \cdot (S(\epsilon/2) - S(\epsilon/4))
\cdot \frac{r}{2}.
\end{eqnarray*}
This completes our proof.

We define $T_1, \ldots, T_{r/2}$ in an inductive fashion. For $i \in
\{0, \ldots, \frac{r}{2}-1 \}$, assume that we have defined $T_1,
\ldots, T_i$ and we intend to define $T_{i+1}$.
Let $\Good_1$ be the set of strings $t \in \{0,1\}^i$ such that
$\Pr(T_1, \ldots, T_i = t) > 2^{-r}$. Then, $$\Pr(T_1, \ldots, T_i \in
\Good_1) \quad \geq \quad 1 - 2^{-r+i} \quad \geq \quad 1 -
2^{-r/2-1}.$$ Let $\epsilon_t$ be 
the error of the protocol $\mcP$ conditioned on $T_1, \ldots, T_i =
t$. Note that $\av[\epsilon_t] $ is the same as the overall expected
error of $\mcP$; hence $\av[\epsilon_t] \leq \epsilon^3/4$. Now using
Markov's inequality we get a set $\Good_2 \in
\{0,1\}^i$ such that $\Pr (T_1 \ldots T_i \in \Good_2)
\geq 1 - \epsilon$ and $\forall t \in \Good_2, \ \epsilon_t 
\leq \epsilon^2/4$. Let $\Good
\defeq \Good_1 \cap \Good_2$. Therefore  (since $r/2 > \log
(1/\epsilon)$, from the hypothesis of the theorem),
\begin{equation} \label{eq:largegood} \Pr(T_1 \ldots T_i \in \Good)
\quad \geq \quad 1 - 2^{-r/2-1} - \epsilon \quad \geq \quad 1 - 2\epsilon.
\end{equation}
For $t \in \{0,1\}^i$ and $y \in \mcY$, let 
$$\delta_{t,y} \defeq
\min\left\{\Pr[f(X,y)=0\vert (T_1 \ldots T_i = t)], \ \Pr[f(X,y)=1\vert (T_1 \ldots
T_i = t)]\right\}.$$ 
Also let, $\epsilon_{t,y}$ be the expected error of $\mcP$
conditioned on $Y=y$ and $T_1 \ldots T_i = t$.

For $t \notin \Good$, we define $T_{i+1} \vert (T_1 \ldots T_i = t) =
0$. Let $t \in \Good$ from now on. Our intention is to identify a
$y_t \in \mcY$, such that $\epsilon_{t,y_t} \leq \epsilon/4$
and $\delta_{t,y_t} \geq
\epsilon/2$. We will then let $T_{i+1} \vert (T_1
\ldots T_i = t)$ to be $f(X,y_t) \vert (T_1
\ldots T_i = t)$.  Lem.~\ref{lem:largeinf} will now imply,
$I(M:T_{i+1} \vert (T_1 \ldots T_i = t)) \geq 
S(\epsilon/2) - S(\epsilon/4)$. Therefore,
\begin{eqnarray*}
I(M:T_{i+1} \vert T_1 \ldots T_i) & \geq & \sum_{t \in \Good} \Pr(T_1 \ldots
T_i = t) \cdot I(M:T_{i+1} \vert (T_1 \ldots T_i = t)) \\
& \geq & (1 - 2\epsilon) \cdot (S(\epsilon/2) - S(\epsilon/4)) \quad
(\text{using  Eq.~\ref{eq:largegood}})
\end{eqnarray*}
and we would be done. 

Now in order to identify a desired $y_t$, we proceed as follows.
Since $r \leq \rec_\epsilon^{1,\mu}(f)$; from the definition of rectangle
bound and given that $\mu$ is a product distribution we have the
following. For all $S \subseteq \mcX$ with $\mu(S \times \mcY) >
2^{-r}$ or in other words with $\Pr[X \in S] > 2^{-r}$,

\begin{equation}
\label{eq:recdef}
\av_{y\leftarrow Y}
\big[\min\left\{\Pr[f(X,y)=0\vert X\in S], \
\Pr[f(X,y)=1\vert X\in S]\right\}\big] \quad > \quad \epsilon.
\end{equation} 

Note that since $t \in \Good$, $\Pr[T_1 \ldots T_i = t] >
2^{-r}$. Hence (\ref{eq:recdef}) implies that $\av_{y\leftarrow
Y}[\delta_{t,y}] > \epsilon$. Now using Markov's inequality and the
fact that, $\forall (t,y),
\delta_{t,y} \leq 1/2$, we get a set $\Good_t \subseteq \mcY$ such that
$\Pr[Y \in \Good_t] \geq \epsilon$ and $\forall y \in \Good_t$, $
\delta_{t,y} \geq \epsilon/2$.

Since $t \in \Good$, we have $\epsilon_t \leq \epsilon^2/4$.  Note
that $\epsilon_t = \av_{y \leftarrow Y}[\epsilon_{t,y}]$. Using a Markov
argument again we finally get a $y_t \in \Good_t$, such that
$\epsilon_{t,y_t} \leq \epsilon/4$. Note that since $y_t \in
\Good_t$, we have $\delta_{t,y_t}
\geq \epsilon/2$ and we are done.

\vspace{0.1in}
\noindent {\bf For total non-boolean functions:}
Let $f: \mcX \times \mcY \rightarrow \mcZ$ be a total non-boolean
function and let $r$ be as before.
We follow the same inductive argument as before to define $T_1
\ldots T_{r/2}$. For $i \in \{0, \ldots, \frac{r}{2}-1\}$, assume that
we have defined $T_1
\ldots T_i$. As before we identify a set $\Good \subseteq \{0,1\}^i$ with $\Pr[T_1
\ldots T_i \in \Good] \geq 1 - 2\epsilon$, such that $\forall t \in \Good, \Pr[T_1
\ldots T_i =t] > 2^{-r}$ and $\epsilon_t \leq \epsilon^2/8$. Since $r \leq
\rec_\epsilon^{1,\mu}(f) $, from the definition of rectangle 
bound and the fact that $\mu$ is product, we have , $\forall S
\subseteq \mcX$ with $\mu(S \times \mcY) > 2^{-r}$,
\begin{equation}
\label{eq:recdef1}
\av_{y\leftarrow Y}
\big[\max_{z \in \mcZ} \left\{\Pr[f(X,y)=z\vert X\in S] \right\}\big]
\quad < \quad 1 - \epsilon.
\end{equation} 
For $t \in \{0,1\}^i$ and $y \in \mcY$, let $\epsilon_{t,y}$ be as
before and let, 
$$\delta_{t,y} \defeq \max_{z \in \mcZ} \left\{\Pr[f(X,y)=z
\vert (T_1\ldots T_i =t)] \right\}.$$
For $t \notin \Good$, let us define $T_{i+1}\vert (T_1\ldots T_i =t)$ to
be $0$.  Let $t \in \Good$ from now on. Note that~(\ref{eq:recdef1})
implies $ \av_{y \leftarrow Y}[\delta_{t,y}] < 1 - \epsilon$. Using
Markov's inequality we get a set $\Good_t
\subseteq \mcY$ with $\Pr[Y \in \Good_t] \geq \epsilon/2$ and $\forall
y \in \Good_t$, $ \delta_{t,y} 
\leq 1 - \epsilon/2$.  Since $\av_{y \leftarrow Y}[\epsilon_{t,y}]=
\epsilon_t \leq \epsilon^2/8$, again 
using a Markov argument we get a $y_t \in \Good_t$, such that
$\epsilon_{t,y_t} \leq \epsilon/4$. Since $\delta_{t,y_t} \leq 1 -
\epsilon/2$ (and $\epsilon \in (0,1/2)$), observe that there would exist a set
$S_{t,y_t} \subseteq \mcZ$ such that, $$\min \{\Pr[f(X,y_t) \in S_{t,y_t} \vert
(T_1\ldots T_i =t)], \Pr[f(X,y_t) \in \mcZ - S_{t,y_t} \vert (T_1\ldots T_i
=t)] \} \quad \geq \quad \epsilon/2.$$

Let us now define $T_{i+1}\vert (T_1\ldots T_i =t)$ to be $1$
if and only if $f(X,y_t) \in S_{t,y_t}\vert (T_1\ldots T_i =t)$ and
$0$ otherwise.  Note that since $\epsilon_{t,y_t} \leq \epsilon/4$,
conditioned on $T_1\ldots T_i =t$, there exists a measurement on $M$,
that can predict the value of $T_{i+1}$ with success probability at
least $1 - \epsilon/4$.  The rest of the proof follows as before.

\vspace{0.25in}
\noindent {\bf For partial non-boolean functions:} 
Let $f: \mcX \times \mcY \rightarrow \mcZ \cup \{*\}$ be a partial
function and let $r$ be as before. Let $i \in \{0, \ldots, \frac{r}{2}
- 1\}$. We
follow a similar inductive argument as in the case of total non-boolean
functions, except for the definition of $T_{i+1}|(T_1 \ldots T_i =
t)$. As before we identify a set $\Good \subseteq \{0,1\}^i$ with $\Pr[T_1
\ldots T_i \in \Good] \geq 1 - 2\epsilon$, such that $\forall t \in \Good, \Pr[T_1
\ldots T_i =t] > 2^{-r}$ and $\epsilon_t \leq \epsilon^5/(2 \cdot 15^4)$.
Since $r \leq
\rec_\epsilon^{1,\mu}(f) $, from the definition of rectangle 
bound and the fact that $\mu$ is product, we have the following. For
all $S \subseteq \mcX$ with $\mu(S \times \mcY) > 2^{-r}$,
\begin{equation}
\label{eq:recdef2}
\av_{y\leftarrow Y}
\big[\max_{z \in \mcZ} \left\{\Pr[f(X,y)= (z \text{ or } *) \vert X\in S] \right\}\big]
\quad < \quad 1 - \epsilon.
\end{equation} 
 For $t
\in \{0,1\}^i$ and $y \in \mcY$, let $\epsilon_{t,y}$ be as 
before and let $$\delta_{t,y} \defeq \max_{z \in \mcZ}
\left\{\Pr[f(X,y)= (z
\text{ or } *) \vert (T_1\ldots T_i =t)] \right\}.$$ 
For $t \notin \Good$, let us define $T_{i+1}\vert (T_1\ldots T_i =t)$ to
be $0$. Let us assume $t \in \Good$ from now on. Let $\Good_t
\subseteq \mcY$ be such that $\forall y \in \Good_t$, $ \delta_{t,y}
\leq 1 - \epsilon/2$.  Using Markov arguments as before we get a $y_t
\in \Good_t$, such that $\delta_{t,y_t} \leq 1 - 
\epsilon/2$ and $\epsilon_{t,y_t} \leq (\epsilon/15)^4 \defeq
\epsilon'$. Since $ \delta_{t,y_t} \leq 1 - \epsilon/2$ it implies
$\Pr[f(X,y_t) = *] \leq 1 - \epsilon/2$. Observe now that can we get a
set $S_{t,y_t} \subseteq \mcZ$ such that, 
\begin{equation} \label{eq:largeS} \min \{\Pr[f(X,y_t) \in
S_{t,y_t} \vert (T_1\ldots T_i =t)],
\Pr[f(X,y_t) \in \mcZ - S_{t,y_t} \vert (T_1\ldots T_i =t)] \} \quad
\geq \quad \epsilon/6.
\end{equation}
Let $O$ be the output of $\bob$ when $Y=y_t$. All along the arguments
below we condition on $T_1\ldots T_i =t$. Note that since $\bob$
outputs some $z\in \mcZ$ even if $f(x,y) = *$,  
let us assume without loss of generality that $q 
\defeq \Pr[O \in S_{t,y_t}] \geq 1/2$ (otherwise similar arguments
would hold by switching the roles of $S_{t,y_t}$ and $\mcZ - 
S_{t,y_t}$).  Let us define $T_{i+1}$ to be $1$ if $(f(X,y_t) \in
S_{t,y_t} \cup \{*\}) $ and $0$ otherwise. Note that
Eq.~(\ref{eq:largeS}) implies $\Pr[ T_{i+1}=1] \leq 1 - \epsilon/6$. Now,
\begin{eqnarray*}
q  & = &  \Pr[O \in S_{t,y_t} \vert (T_{i+1}=1)]  
 \cdot \Pr[ T_{i+1}=1]  \\
& & + \Pr[O \in S_{t,y_t} \text{ and } T_{i+1}=0 ] \\ 
& \leq  & \Pr[O \in S_{t,y_t} \vert (T_{i+1}=1)] \cdot
\Pr[ T_{i+1}=1]   + \epsilon' \\
& \leq &  \Pr[O \in S_{t,y_t} \vert ( T_{i+1}=1)]
\cdot (1 - \epsilon/6)  + \epsilon'
\end{eqnarray*}
This implies,
\begin{eqnarray*}
\Pr[O \in
S_{t,y_t} \vert (T_{i+1}=1)] & \geq & \frac{q -
\epsilon'}{1 - \epsilon/6}  \\ 
 & \geq & (q
 - \epsilon')(1 + \epsilon/6) \\
& = & q + q\epsilon/6 - \epsilon'(1 + \epsilon/6) \\
& \geq & q + \epsilon/12 - \epsilon(1 + 1/12)/(2^3 \cdot 15^4)
\qquad \text{(since $q \geq 1/2$ and $\epsilon \leq 1/2$)}\\
& \geq & q + 0.08\epsilon 
\end{eqnarray*}
Let us define $O' =1$ iff $O \in S_{t,y_t}$ and $O'=0$ otherwise. Then, 
\begin{eqnarray*}
I(M:T_{i+1})  & \geq & I(O' : T_{i+1}) \\ 
& = & S(O') - \Pr[T_{i+1} = 1] \cdot S(O' \vert (T_{i+1} = 1))  \\ 
& & - \Pr[T_{i+1} = 0] \cdot S(O' \vert (T_{i+1} = 0))  \\
& \geq & S(q) - S(q + 0.08\epsilon) - S(\epsilon') \\
& \geq & 1 - S(0.5 + 0.08\epsilon) - S(\epsilon') \\
& \geq&  1 - (1 - 2(0.08\epsilon)^2) - 2(\epsilon/15)^2 \\
& \geq & \epsilon^2/300
\end{eqnarray*}
The third inequality above follows since the function $S(p)$ is
concave and monotonically decreasing in $[\frac{1}{2},1]$. The fourth
inequality follows from Fact~\ref{fact:ssmall}.  The rest of the proof
follows as before. 
\end{proofof}

\section{Application: Security of boolean extractors against
quantum adversaries}
\label{sec:qextract}
In this section we present a consequence our lower bound result
Thm.~\ref{thm:reclower} to prove security of extractors
against quantum adversaries. In
this section we are only concerned with boolean extractors. We
begin with following definitions.
\begin{definition}[Min-entropy] Let $P$ be a distribution on
$[N]$. The  {\em min-entropy} of $P$ denoted $S_\infty(P)$ is defined
to be $ - \log \max_{i \in [N]} P(i)$.
\end{definition}

\begin{definition} [Strong extractor]  
\label{def:ext}
Let $\epsilon \in (0,1/2)$. Let $Y$ be
uniformly distributed on $\mcY$.  A {\em strong $(k,
\epsilon)$-extractor} is a function $h: \mcX \times \mcY \rightarrow
\{0,1\}$ such that for any random  variable $X$ distributed on $\mcX$
which is independent of $Y$ and with $S_\infty(X) \geq k$ we have,
$$ \| h(X,Y)Y - U \otimes Y \|_1 \quad < \quad 2\epsilon,$$
where $U$ is the uniform distribution on $\{0,1\}$. 

In other words, even given $Y$ (and not $X$);
$h(X,Y)$ is still close (in $\ell_1$ distance) to being a uniform bit.
\end{definition}

Let $X,Y,h$ be as in the definition above.  Let us consider a random
variable $M$, taking values in some set $\mcM$, correlated with $X$ and
independent of $Y$. Let us now limit the correlation that $M$ has with
$X$, in the sense that $\forall m \in \mcM, S_\infty(X \vert
M=m) \geq k$. Since $h$ is a strong $(k,\epsilon)$-extractor, it is
easy to verify that in such a case,
\begin{eqnarray*}
\forall m \in \mcM, \quad \| h(X,Y)Y \vert (M=m) - U \otimes Y \vert
(M=m) \|_1 & <  & 2\epsilon \\ 
\Rightarrow \quad \| h(X,Y)YM - U \otimes YM \|_1 & <  & 2\epsilon 
\end{eqnarray*}
In other words,
still close (in $\ell_1$ distance) to being a uniform bit.

Now let us ask what happens if the system $M$ is
a quantum system. In that case, is it still true that given $M$ and
$Y$, $h(X,Y)$ is close to being a uniform bit? This question has
been increasingly studied in recent times specially for its
applications for example in  privacy amplification in Quantum key
distribution protocols and in the  Quantum bounded
storage models~\cite{KonigMR05,KonigR05,KonigT08}. 

However when $M$ is a quantum system, the min-entropy of $X$,
conditioned on $M$, is not easily captured since conditioning on a quantum
system needs to be carefully defined. An alternate way to capture
the correlation between $X$ and $M$ is via the {\em guessing
probability}. Let us consider the following definition.

\begin{definition}[Guessing-entropy] Let $X$ be a classical random
variable taking values in $\mcX$.  Let $M$ be a correlated
quantum system with the joint classical-quantum state being $\rho_{XM} =
\sum_x \Pr[X=x] \ket{x}\bra{x} \otimes \rho_x$. Then the {\em guessing-entropy}
of $X$ given $M$, denoted $S_g(X \leftarrow M)$ is defined to be:
$$ S_g(X \leftarrow M) \defeq - \log \max_\mcE \sum_x \Pr(X=x) \Tr (E_x
\rho_x) $$
where the maximum is taken over all $\mathsf{POVM}$s
$\mcE \defeq \{E_x : x \in \mcX \}$. (Please refer
to~\cite{NielsenC00} for a definition of $\mathsf{POVM}$s).
\end{definition}
The guessing-entropy turns out to be a useful notion in the quantum
contexts. Let $h,X,Y, M$ be as before, where $M$ is a quantum system.
K\"{o}nig and Terhal~\cite{KonigT08} have (roughly) shown that if the
guessing entropy $S_g(X \leftarrow M)$, is at least $k$, then given
$M$ and $Y$ (and not $X$), $h(X,Y)$ is still close to a uniform
bit. We state their precise result here.

\begin{theorem}
\label{thm:extractkonig}
Let $\epsilon \in (0, 1/2)$. Let $h: \mcX
\times \mcY \rightarrow \{0,1\}$ be a strong
$(k,\epsilon)$-extractor. Let $U$ be the uniform distribution on
$\{0,1\}$. Let $YXM$ be a classical-quantum system with $YX$ being
classical and $M$ quantum. Let $Y$ be uniformly distributed and
independent of $XM$ and, $$ S_g( X \leftarrow M) \quad >
\quad k + \log 1/\epsilon. $$ Then,
$$ \| h(X,Y)YM - U \otimes YM \|_1 \quad
< \quad 6 \sqrt{\epsilon}. $$ 
\end{theorem}

We show a similar result as follows.

\begin{theorem}
\label{thm:extractour}
Let $\epsilon \in (0, 1/2)$. Let $h: \{0,1\}^n
\times \{0,1\}^m \rightarrow \{0,1\}$ be a strong
$(k,\epsilon)$-extractor. Let $U$ be the uniform distribution on
$\{0,1\}$. Let $YXM$ be a classical-quantum system with
$YX$ being classical and $M$ quantum. Let $X$ be uniformly
distributed on $\{0,1\}^n$. Let $Y$ be uniformly distributed on
$\{0,1\}^m$ and independent of $XM$ and,
\begin{equation} \label{eq:highinf} I(X:M) \quad < \quad
b(\epsilon) \cdot (n-k).
\end{equation}
Then,
\begin{equation} \label{eq:lowerr} \| h(X,Y)YM - U \otimes YM \|_1
\quad < \quad 1 - a(\epsilon)  
\end{equation}
where $a(\epsilon) \defeq \frac{1}{4} \cdot (\frac{1}{2} -
\epsilon)^3$ and $b(\epsilon) \defeq \epsilon \cdot (S(\frac{1}{4} -
\frac{\epsilon}{2}) - S(\frac{1}{8} - \frac{\epsilon}{4})). $
\end{theorem}
Before proving Thm.~\ref{thm:extractour}, we will make a few points
comparing it with Thm.~\ref{thm:extractkonig}. 
\begin{enumerate}
\item 
Let's observe that if $M$ is a classical system, then 
\begin{eqnarray*}
S_g(X \leftarrow M) & = &  - \log \av_{m \leftarrow M} [2^{-S_\infty(X \vert
M=m)}] \\
& \leq &  \av_{m \leftarrow M} [S_\infty(X \vert M=m) \cdot \log_e 2] \\
& \leq & \av_{m \leftarrow M} [S_\infty(X \vert M=m) ] \\
& \leq & S(X \vert M)
\end{eqnarray*}
The first inequality follows from the convexity of the exponential
function. The last inequality follows easily from definitions.
This implies, 
\begin{equation} \label{eq:infless} 
I(X:M) \quad = \quad S(X) - S(X \vert M) \quad \leq \quad S(X) -
S_g(X \leftarrow M).
\end{equation}
So if $M$ is classical, then the implication of
Thm.~\ref{thm:extractour} appears stronger than the implication in
Thm.~\ref{thm:extractkonig} (although being weak in terms of the
dependence on $\epsilon$.) We cannot show the inequality
(\ref{eq:infless}) when $M$ is a quantum system but conjecture it to be
true. If the conjecture is true,
Thm.~\ref{thm:extractour} would have stronger implication than
Thm.~\ref{thm:extractkonig} in the quantum case as well.

\item The proof of Thm.~\ref{thm:extractkonig} in~\cite{KonigT08}
crucially uses some properties of the so called {\em pretty good
measurements} ($\mathsf{PGM}s$). Our result follows here without
using $\mathsf{PGM}$s and via completely different arguments.

\item Often in applications concerning the Quantum bounded storage model, an upper bound
on the number of qubits of $M$ is available. This implies the same upper
bound on $I(X:M)$. If this bound is
sufficiently small such that it suffices the assumption of 
Thm.~\ref{thm:extractour}, then $h$ could be used to extract a private
bit successfully, in the presence of a quantum adversary.
\end{enumerate}

Let us return to the proof of Thm.~\ref{thm:extractour}. 
We begin with the following  key observation. It
essentially states that a boolean function which can extract a bit
from sources of low min-entropy has high one-way rectangle bound
under the uniform distribution.

\begin{lemma}
\label{lem:largerec}
Let $\epsilon \in (0, 1/2)$. Let $h:
\{0,1\}^n \times \{0,1\}^m \rightarrow \{0,1\}$ be a strong
$(k,\epsilon)$-extractor. Let $\mu \defeq U_n \otimes U_m$, where
$U_n, U_m$ 
are uniform distributions on $ \{0,1\}^n$ and $\{0,1\}^m$
respectively. Then $$ \rec^{1,\mu}_{1/2 -
\epsilon}(h) \quad > \quad n - k.$$
\end{lemma}
\begin{proof}
Let $R \defeq S \times \{0,1\}^m$  be any one-way rectangle where $S
\subseteq \{0,1\}^n$ with $\mu(R) \geq 2^{-n+k}$ which essentially
means that $|S| \geq 2^{k}$. Let $X$ be uniformly
distributed on $S$. This implies that $S_\infty(X) \geq k$. Let $Y$ be
uniformly distributed on $\{0,1\}^m$.  Since $h$ is a strong
extractor, from Definition~\ref{def:ext} we have (where $U$ is the
uniform distribution on $\{0,1\}$):
\begin{eqnarray*}
  \| h(X,Y)Y - U \otimes Y \|_1 & < & 2 \epsilon \\
 \Leftrightarrow \av_{y \leftarrow Y} [\| h(X,y) - U \|_1] & < &  2\epsilon
\end{eqnarray*}
  We note that from 
  Definition~\ref{def:mono}, above implies that $\mu_R$ is not
  $1/2 + \epsilon$ monochromatic. Hence from the definition of
  the rectangle bound, Definition~\ref{def:rec} we have $\rec^{1,\mu}_{1/2 -
  \epsilon}(h) > n - k$.
\end{proof}

We will also need the following information theoretic fact.
\begin{lemma}
\label{lem:helstrom}
Let $RQ$ be a joint classical-quantum system where $R$ is a classical
boolean random variable.  For $a \in \{0,1\}$, let the quantum state
of $Q$ when $R=a$ be $\rho_a$.  Then there is a measurement that can be
done on $Q$ to guess value of $R$ with probability $\frac{1}{2} +
\frac{1}{2} \cdot \|RQ - U \otimes Q \|_1$.

\begin{proof}
Let us note that $$ \|RQ - U \otimes Q \|_1 = \| \Pr[R=0]\rho_0 -
\Pr[R=1]\rho_1 \|_1.$$
Now Helstrom's Theorem (Thm.~\ref{thm:helstrom}) immediately helps us
conclude the desired.
\end{proof}
\end{lemma}

We are now ready for the proof of Thm.~\ref{thm:extractour}.

\begin{proofof}{Thm.~\ref{thm:extractour}}
 We prove our result in the contrapositive manner. Let,
\begin{equation*}   \| h(X,Y)MY
- U \otimes MY \|_1 \quad > \quad 1 - a(\epsilon).   \end{equation*}
Note that this is equivalent to:
\begin{equation} \label{eq:lowerr1}  \av_{y \leftarrow Y} [\| h(X,y)M
- U \otimes M \|_1] \quad > \quad 1 - a(\epsilon).   \end{equation}

Let's consider a one-way communication protocol $\mcP$ for $h$ where the
 inputs $X$ and $Y$ of $\alice$ and $\bob$ respectively are drawn
 independently from the uniform distributions on $\{0,1\}^n$ and
 $\{0,1\}^m$ respectively. Let $\mu$ be the distribution of $XY$. 
Now let $M$ be sent as the message of $\alice$ in $\mcP$. Note that now
(\ref{eq:lowerr1}) along with Lem.~\ref{lem:helstrom} implies that
the distributional error of $\mcP$ will be at most $a(\epsilon)/2 =
\frac{1}{8} \cdot (\frac{1}{2} - \epsilon)^3$. Let $\epsilon' \defeq
1/2 - \epsilon$. Therefore $\mcP$ has distributional error at most $\epsilon'^3/8$.
Arguing as in the proof of Thm.~\ref{thm:reclower} we get that,
\begin{eqnarray*}
I(X:M) & \geq & \frac{1}{2} \cdot (1 - 2\epsilon') (S(\epsilon'/2) -
S(\epsilon'/4)) \cdot \rec_{\epsilon'}^{1,\mu}(h) \\
& = & \epsilon \cdot (S(\frac{1}{4} -
\frac{\epsilon}{2}) - S(\frac{1}{8} - \frac{\epsilon}{4})) \cdot \rec_{1/2 -
\epsilon}^{1,\mu}(h) \\
& = & b(\epsilon) \cdot \rec_{1/2 -
\epsilon}^{1,\mu}(h) \\
& > & b(\epsilon)\cdot (n-k)
\end{eqnarray*}
The last inequality follows from Lem.~\ref{lem:largerec} since $h$ is
a strong $(k,\epsilon)$-extractor.

\end{proofof}

\section{Conclusion}
\label{sec:conclude}
In the wake of our quantum lower bound result, it is natural to ask 
whether in the two-way model also, there is a similar relationship between 
quantum distributional communication complexity of a function $f$,
under product distributions, and the corresponding rectangle bound.

Concerning the classical upper bound, a natural question to ask is
whether the bound could be tightened, specially in terms of its
dependence on the mutual information $I(X:Y)$ between the inputs, under
a given non-product distribution? 
For example, could it be 
that for a boolean function $f$ and a distribution $\mu$ on the
inputs, $\sD_{\epsilon}^{1,\mu}(f) =
O(I(X:Y) + \vc(f))$?

\subsection*{Acknowledgment} We thank Ashwin Nayak for many helpful discussions.

\bibliography{one-way}


\appendix

\section{}
\label{sec:example}
Let $ n \geq 1$ be a sufficiently large
integer. Let the Noisy Partial Matching $\mathsf {(NPM_n)}$ function be 
as follows. 

\begin{center}
\fbox{
\begin{minipage}[1]{5.25in}
{\bf Input:} 

$\alice$: A string $x \in \{0,1\}^n$. 

$\bob$: A string $w \in \{0,1\}^n$ and a Matching $M$ on $[2n]$ comprising of
$n$ disjoint edges. 
\\ \\
{\bf Output:} 

For a matching $M$ and a string $x$, let $Mx$ represent the $n$ bit
string corresponding to the $n$ edges of $M$ obtained as follows. For
an edge $e \defeq (i,j)$ in $M$ the bit included in $Mx$ is $x_i
\oplus x_j$, where $x_i, x_j$ represent the $i,j$-th bit of $x$.
\\ \\
Output bit $b \in \{0,1\}$ if and
only if the Hamming distance between strings $(Mx) \oplus b^n$ and $w$
is at most n/3.  If there is no such bit $b$ then output $0$.
\end{minipage}
}
\end{center}

Now let the non-product distribution $\mu$ on inputs of $\alice$ and
$\bob$ be as follows. Let $\alice$ be given $x$ drawn uniformly from
$\{0,1\}^n$. Let $\bob$ be given matching $M$ drawn  uniformly from
the set of all matchings on $[2n]$. With probability $1/2$, $\bob$ is
given $w$  uniformly from the set of all strings with Hamming distance
at most $n/3$ from $Mx$ and with probability $1/2$, he is given $w$
uniformly from the set of all strings with Hamming distance
at most $n/3$ from $(Mx) \oplus 1^n$. Note that in $\mu$ there is
correlation between the inputs of $\alice$ and $\bob$ and hence $\mu$
is non-product. Now we have the following.
\begin{theorem}[\cite{Gavinskyetal07}, implicit] 
Let $n \geq 1$ be a sufficiently large integer and let $\epsilon \in (0,1/2)$. Let
$\mathsf{NPM}_n$ and $\mu$ be as described above. Then,
$\rec^{1,\mu}_{\epsilon}(\mathsf{NPM}_n) = \Omega(\sqrt{n})$ whereas
$\sQ^{1,\mu}_\epsilon(\mathsf{NPM}_n) = O(\log n)$. 
\end{theorem}

\end{document}